\documentclass[twocolumn]{aastex61}
\usepackage{amsmath,amstext}
\usepackage[figure,figure*]{hypcap}
\usepackage{natbib}
\usepackage{xcolor}

\newcommand{\beq} {\begin{equation}}
\newcommand{\cs} {c_{\rm s}}
\newcommand{\dotEg} {\dot{E}_{\rm g}}
\newcommand{\dotEt} {\dot{E}_{\rm turb}}
\newcommand{\eeq} {\end{equation}}
\newcommand{\Eg} {E_{\rm g}}

\newcommand{\etaml} {\eta_{\rm ML}}
\newcommand{\etasim} {\eta_{\rm sim}}
\newcommand{\Eturb} {E_{\rm turb}}

\newcommand{\fsim} {f_{\rm sim}}

\newcommand{\gamef} {\gamma_{\rm e}}
\newcommand{\ggeq} {g_{\rm eq}}
\newcommand{\gsim} {g_{\rm sim}}
\newcommand{\gvir} {g_{\rm vir}}
\newcommand{\hsim} {h_{\rm sim}}
\newcommand{\Lb} {L_{\rm b}}
\newcommand{\Ld} {{\cal L}_{\rm d}}
\newcommand{\LJ} {L_{\rm J}}

\newcommand{\Pram}{P_{\rm r}}
\newcommand{\Pt}{P_{\rm t}}

\newcommand{\sigod}{\sigma_{\rm 1D}}
\newcommand{\sigtd}{\sigma_{\rm 3D}}
\newcommand{\sigvr}{\langle v_{\rm rad}^2 \rangle}
\newcommand{\tff} {t_{\rm ff}}
\newcommand{\vg} {v_{\rm g}}
\newcommand{\vphi} {v_\varphi}
\newcommand{\vrad} {v_{\rm rad}}
\newcommand{\vradtovg} {\frac{\langle \vrad \rangle} {\vg}}

\newcommand{\vtan} {v_{\rm tan}}
\newcommand{\vthet} {v_\theta}

\shorttitle{Non-adiabatic turbulence driving during gravitational collapse} 
\shortauthors{Guerrero-Gamboa and V\'azquez-Semadeni}

\begin{document}

\title{Non-adiabatic turbulence driving during gravitational collapse}
\author{Rub\'en Guerrero-Gamboa}
\author{Enrique V\'azquez-Semadeni} 
\affiliation{Instituto de Radioastronom\'ia y Astrof\'isica, Universidad Nacional Aut\'onoma de M\'exico, Campus Morelia Apartado Postal 3-72, 58090 Morelia, Michoac\'an, M\'exico}

\begin{abstract}

We investigate the generation of turbulence during the prestellar gravitational contraction of a turbulent spherical core. We define the ratio $g$ of the one-dimensional turbulent velocity dispersion, $\sigod$ to the gravitational velocity $\vg$, to then analytically estimate $g$ under the assumptions of a) equipartition or virial equilibrium between the gravitational ($\Eg$) and turbulent kinetic ($\Eturb$) energies and b) stationarity of transfer from gravitational to turbulent energy (implying $\Eturb/\Eg=$cst). In the equipartition and virial cases, we find $g=\sqrt{1/3}\approx0.58$ and $g=\sqrt{1/6}\approx0.41$, respectively; in the stationary case we find $g=\langle\vrad\rangle\Ld/(4\pi\sqrt{3}\eta R\vg)$, where $\eta$ is an efficiency factor, $\Ld$ is the energy injection scale of the turbulence, and $R$ is the core's radius. Next, we perform AMR simulations of the prestellar collapse of an isothermal, transonic turbulent core at two different resolutions, and a non-turbulent control simulation. We find that the turbulent simulations collapse at the same rate as the non-turbulent one, so that the turbulence generation does not significantly slow down the collapse. We also find that a) the simulations approach near balance between the rates of energy injection from the collapse and of turbulence dissipation; b) $g\approx0.395\pm0.035$, close to the ``virial'' value (turbulence is $\sim35-40\%$ of non-thermal linewidth); c) the injection scale is $\Ld\lesssim R$, and d) the ``turbulent pressure'' $\rho\sigod^2$ scales as $\sim\rho^{1.64}$, an apparently nearly-adiabatic scaling. We propose that this scaling and the nearly virial values of the turbulent velocity dispersion may be reconciled with the non-delayed collapse rate if the turbulence is dissipated as soon as it is generated.

\end{abstract}

\keywords{gravitation --- hydrodynamics --- ISM: clouds --- turbulence}

\section{Introduction}

Turbulence in molecular clouds (MCs) and their substructures (clumps, filaments and cores) is generally thought to play a role of support against their self-gravity, which can prevent or delay the collapse of such structures, until it is dissipated, at which time a dense core can lose support and proceed to collapse \citep[e.g.] [] {MK04, BT07, MO07, HF12}. However, the origin, and therefore the fate of such turbulence remains unclear, and in fact, it is unclear whether the observed nonthermal motions correspond to mostly to infall or to true turbulence that can provide a ram pressure capable of opposing the collapse, or to a combination of both in an unknown proportion. 

The fact that the nonthermal motions appear to be close to virialization at all scales in MCs and their substructures \citep[e.g.,] [] {Heyer+09, BP+11, Traficante+18} suggests that the origin of these motions may be related to, or driven by, gravity \citep{VS+07, BP+11}. In addition, it has been argued by various numerical and analytical studies \citep[e.g.,] [] {VS+98, Field+08, KH10, Sur+10, Federrath+11, RG12, MC15, Murray+17, Li18} that the turbulence itself (i.e., not the infall motions) is driven by the collapse, producing a situation analogous to adiabatic heating by compression, and so this mechanism has been sometimes loosely termed ``adiabatic heating'' of the turbulence \citep[e.g.,] [] {RG12, MC15, Murray+17}. It is important to note, however, that the analogy can, at best, only be partial, because turbulence is an intrinsically dissipative phenomenon, while adiabaticity in the thermodynamic sense implies no heat losses. So, if the turbulence is driven (``heated'') by the collapse-induced compression, at most it must be so in a lossy, rather than adiabatic, way. 

The extent to which the collapse can drive the turbulence is crucial in determining whether collapse-driven turbulence can eventually slow down the collapse, and there are conflicting results in the literature. For example, while \citet{Murray+17} found that the turbulence generated during the collapse is capable of significantly slowing the latter, \citet{Sur+10} and \citet{Federrath+11} found that the infall speed is essentially equal to the free-fall speed at the boundary of the cores whose collapse they simulated, implying that no slow-down of the collapse occurred. \citet{Murray+17} attributed the discrepancy to the absence of external driving to the turbulence. \citet{Li18} concluded, from order-of-magnitude considerations, that the dissipation of turbulence is slow enough as to cause the effective infall speed to be as small as 20-50\% of the free-fall value. 

\citet[] [hereafter RG12] {RG12} employed a novel technique that considered a contracting reference frame (analogous but opposite to cosmological prescriptions for the expanding Universe) for both their analytical and their numerical calculations. Within this framework, they wrote an ``adiabatic heating'' equation for the system, and considered cases where the initial turbulent turnover rate (or energy cascade, turbulent, or dissipation rate, proportional to the inverse of the turbulent crossing time) was either smaller or larger than the system's contraction rate (or the collapse rate, or roughly the inverse of the free-fall time, for a gravitationally-contracting system).\footnote{RG12 actually discussed in terms of the ``eddy turnover frequency'' of the turbulence and the ``Hubble frequency'' of their domain. We find it more intuitive here to discuss in terms of the turbulent dissipation rate and the collapse rate, respectively.} They found that, when the turbulent dissipation rate was smaller than the contraction rate, then the former increased, being ``heated'' by the collapse. Conversely, they found that, when the initial turbulent rate was larger than the contraction rate, the former {\it decreased}. Thus, RG12 concluded that the system should evolve toward equating the two rates. This result suggests that the system should evolve toward establishing either equipartition between the turbulent kinetic and gravitational energies, or more generally to a stationary state in which the energy ratio where the turbulent transfer and collapse rates become equal, although not necessarily implying energy equipartition. Equipartition would be the expected state in the case of dissipationless free-fall collapse. More recently, \citet{Xu+20} have performed a similarity analysis of the problem, and one of their findings is that the infall speed is unperturbed when the injection and dissipation rates balance each other.

In this paper we present in Sec.\ \ref{sec:analyt_pred} a simple analytical study to determine the expected ratio $g \equiv \sigod/\vg$ of the one-dimensional turbulent velocity dispersion $\sigod$ to the gravitational velocity, $\vg$, under two plausible assumptions: i) energy equipartition or ii) a stationary state where the ratio of the turbulent to infall kinetic energies remains constant. In Sec.\ \ref{sec:sims} we then present a numerical simulation of the process of collapse in the presence of initial transonic turbulence, to numerically determine which of the two regimes is actually realized, and the value of the ratio $g$ (Sec.\ \ref{sec:results}). In Sec.\ \ref{sec:disc} we discuss our results in the context of previous work, and finally, in Sec.\ \ref{sec:concls}, we present a summary and our conclusions. Note that in the present study we have chosen to restrict our analysis to the HD case, since most of the recent works investigating the generation of turbulence by the collapse, which our study extends, also have been restricted to the HD case \citep[][although the latter authors do devote a section to discuss the likely enhancement of reconnection diffusion due to the generation of turbulence by the collapse]{RG12,MC15,Li18,Xu+20}.

\section{Analytical estimates} \label{sec:analyt_pred}

\subsection{Definitions} \label{sec:defs}

In this section, we investigate the generation of turbulence during spherical gravitational collapse analytically, assuming that this generation mechanism achieves a stationary regime. Specifically, we present an analytical estimation of the expected value of the ratio $g \equiv \sigod/\vg$ of the one-dimensional turbulent velocity dispersion, $\sigod$, to the gravitational velocity given by
\begin{equation}
    \vg = \sqrt{\frac{2 \beta GM} {R}}, 
    \label{eq:vg}
\end{equation}
of a spheric cloud of mass $M$ and radius $R$. Here, $\beta$ is a geometrical factor of order unity. We perform the estimate under two alternative reasonable assumptions: i) that the system approaches equipartition or virial balance between the gravitational ($\Eg$) and the turbulent kinetic energy ($\Eturb$), or ii) that the system attains an arbitrary, yet self-consistent, stationary value of the ratio of the two energies, where the energy injection rate balances the dissipation rate.

We start by decomposing the total velocity in spherical coordinates into its radial and tangential components, $\vrad$ and $\vtan$, where 
\begin{equation}
    \langle \vtan^2 \rangle = \frac{\langle v_\theta^2 \rangle + \langle v_\varphi^2\rangle} {2} \equiv \sigod^2, 
    \label{eq:vtan}
\end{equation}
where the last equality states that we identify $\langle \vtan^2 \rangle^{1/2}$ with the one-dimensional velocity dispersion of the turbulence. We use this definition instead of the usual definition $\sigod^2 = \left(\langle \vrad^2 \rangle + \langle \vphi^2 \rangle + \langle \vthet^2 \rangle\right)/3$ because $\vrad$ contains a possibly dominant contribution from infall rather than from turbulence. On the other hand, note that $\langle \vrad^2 \rangle^{1/2}$ is expected to contain contributions from both the turbulent fluctuations and the infall speed, so that  
\beq
    \langle \vrad^2 \rangle = \sigod^2 + \vg^2.
    \label{eq:vdisp_rad}
\eeq
We thus define the ratio 
\begin{equation}
    f^2 \equiv \frac{\langle \vrad^2 \rangle}  {\vg^2}
    \label{eq:def_f}
\end{equation}
as a measure of the excess kinetic energy in the collapse due to the turbulence, which in Sec.\ \ref{sec:measurements} we will measure in our numerical simulations at one particular radius.

The main quantity we are interested in is the ratio of the one-dimensional turbulent velocity dispersion $\sigod$ to the gravitational speed,
\begin{equation}
    g^2 \equiv \frac{\sigod^2} {\vg^2}, 
    \label{eq:def_g}
\end{equation}
which determines the fraction of energy going from the gravitational contraction to the turbulent motions.

From eqs.\ (\ref{eq:def_f}) and (\ref{eq:def_g}), we also obtain
\beq
    \frac{\sigod^2} {\sigvr} = \frac{g^2} {f^2} \equiv h^2,
    \label{eq:sig_vrad_ratio}
\eeq
which is another quantity we will measure directly from our numerical simulations, in order to obtain $g$.

\subsection{Energy equipartition and virialization} 
\label{sec: equip}

We now estimate the value of the ratio $g$ in the case that the infall kinetic energy attains equipartition or virial equilibrium with the gravitational energy. First, assuming that the turbulence is isotropic, we write the turbulent kinetic energy as
\begin{equation}
    \Eturb \approx \frac{1}{2}\sigtd^2 M = \frac{3}{2} \sigod^2 M. 
    \label{eq:Eturb}
\end{equation}
Inserting eq.\ (\ref{eq:def_g}) in (\ref{eq:Eturb}), we thus find
\begin{equation}
    \Eturb = \frac{3}{2} g^2 \vg^2 M.
    \label{eq:Eturb_v2}
\end{equation}
On the other hand, we have, for the gravitational energy,
\begin{equation}
    \Eg = - \frac{1}{2} \vg^2 M.
    \label{eq:Eg}
\end{equation}
Therefore, in the case of equipartition, where $\Eturb = |\Eg|$, we find
\begin{equation}
    \ggeq = \sqrt{\frac{1} {3}} \approx 0.58.
    \label{eq:g_equip}
\end{equation}
where we have denoted by $\ggeq$ the value of $g$ corresponding to equipartition.

In the case of virial equilibrium, in which $2 \Eturb = |\Eg|$, we obtain
\begin{equation}
    \gvir = \sqrt{\frac{1}{6}} \approx 0.41.
    \label{eq:g_vir}
\end{equation}
\subsection{Self-consistent stationary regime} \label{sec:stat_reg}

We now consider the possibly more realistic situation that the system adjusts itself to a stationary regime in which the rate of energy injection into the turbulence due to the release of gravitational energy balances the dissipation rate of the turbulence by viscosity. We write this condition as
\begin{equation}
    {\dot E}_{\rm turb} = {\dot E}_{\rm g}.
    \label{eq:stat_cond}
\end{equation}
This condition has been investigated by \citet{Li18}, although this author made some {\it ad hoc} assumptions about the parameters entering the above rates, while instead here we leave them open, to be determined from the results of our numerical simulations.

To compute the energy injection rate from the release of gravitational energy, we differentiate the gravitational energy given by eqs.\
(\ref{eq:vg}) and (\ref{eq:Eg})
\begin{eqnarray}
    \dotEg &=& \frac{d} {dt} \left(- \frac{\beta GM^2} {R} \right) \approx \frac{\beta G M^2 \langle \vrad \rangle} {R^2} \nonumber \\
           &=& - \vradtovg \left( \frac{2|\Eg|^3} {M R^2}\right)^{1/2}, 
    \label{eq:dotEg}
\end{eqnarray}
where in the first equality we have written $\dot R \approx \langle \vrad \rangle$, and in the second equality we have introduced the velocity $\vg$ as defined by eq.\ (\ref{eq:vg}). Note that the ratio $\langle \vrad \rangle/\vg$ is not the same as the ratio $f$ defined in eq.\ (\ref{eq:def_f}), which involves the variance of $\vrad$, while $\langle \vrad \rangle/\vg$ involves the mean value of $\vrad$.

On the other hand, for the kinetic energy dissipation rate, we use the expression provided by \citet[] [hereafter, ML99] {ML99}: 
\begin{equation}
    \dotEt = - \frac{2 \pi \etaml M \sigtd^3} {\Ld},
    \label{eq:dotEk1}
\end{equation}
where $M$ is the mass of the system, $\Ld$ is the energy-injection scale of the turbulence, and $\etaml \approx 0.21/\pi \approx 0.067$ is a constant he determined from numerical simulation of driven isothermal turbulence in a periodic box. We refer in general to the proportionality constant, $\eta$, as the {\it dissipation efficiency}, and allow for the possibility that it differs in our problem from the value found by ML99, $\etaml$, due to the different nature of the energy injection, so we drop the subindex ``ML''. Thus, inserting eq.\ (\ref{eq:Eturb}) into eq.\ (\ref{eq:dotEk1}), we obtain
\begin{equation}
    \dotEt = - 2 \pi \eta \left(\frac{2^3 \Eturb^3} {M \Ld^2} \right)^{1/2}.
    \label{eq:dotEk2}
\end{equation}
We can now compute the derivative of the ratio of the two energies (in absolute value, since we equate the turbulent dissipation rate to the gravitational injection rate), and set it to zero to demand
stationarity:
\[
    \frac{d}{dt}\left(\frac{\Eturb} {|\Eg|}\right) = \frac{|\Eg| |\dotEt| - |\Eturb| |\dotEg|} {\Eg^2}  = 0. 
\]
This condition is satisfied when
\[
    4 \pi \eta \frac{|\Eg| \Eturb^{3/2}} {\Ld} = \vradtovg \frac{\Eturb |\Eg|^{3/2}} {R},
\]
which implies
\begin{equation}
    \Eturb = \left(\frac{1} {4 \pi \eta} \vradtovg \frac{\Ld} {R} \right)^{2} |\Eg|.
    \label{eq:stat_cond2}
\end{equation}
This implies that the velocity ratio $g$ in the stationary regime is
\begin{equation}
    g_{\rm st} = \frac{1} {4 \pi \sqrt{3}} \vradtovg \frac{\Ld} {\eta R}.
    \label{eq:g_stat}
\end{equation}

If we adopt the value of $\eta$ found by ML99, we find
\begin{equation}
    g_{\rm st,ML} \approx 0.687 \vradtovg \frac{\Ld} {R}.
    \label{eq:g_stat_ML}
\end{equation}
In the next section, we will estimate the ratio $\langle \vrad \rangle/\vg$ from our numerical simulations. This will allow calculation of the ratio of the driving scale to the core's radius, {\it provided we assume $\eta = \etaml$}. Otherwise, eq.\ (\ref{eq:g_stat}) shows that there is an unresolvable degeneracy between the dissipation efficiency and the ratio of the driving scale to the radius of the sphere, which in our case is not known.  It can be {\it assumed} that it is the radius or the diameter of the core \citep[as done by] [] {Li18}, but our numerical experiments will not allow us to resolve this degeneracy.

\section{The simulations} \label{sec:sims}

We use the Eulerian adaptive mesh refinement (AMR) code \textsc{FLASH}2.5 \citep{Fryxell+00} to perform two 3D numerical simulations of the gravitational collapse of a dense core in presence of a turbulent velocity field, to determine the transfer of energy from the gravitational to the turbulent motions. These two turbulent simulations differ in the maximum resolution and refinement criterion, to test for convergence. We also consider a reference non-turbulent simulation. 

We consider a numerical box filled with isothermal gas, so that the simulation is scale-free. The sound speed and the mean density are set so that the box size is $\Lb \approx 2.48 \LJ$, where $\LJ = (\pi \cs^2/G \langle \rho \rangle)^{1/2}$ is the Jeans length, and $\langle \rho \rangle$ is the mean density. The boundary conditions are perdiodic for the hydrodynamics, and isolated for the self-gravity.

The simulations are initiated with a uniform density $\rho_0$, on top of which a spherically symmetric Gaussian profile, centered at the center of the numerical box, is added. This Gaussian profile has a peak value of $\rho_{\rm max} = 2.5 \rho_0$ and a width $\sigma_\rho = 0.25 \Lb$. Therefore, the maximum density within the box is $\rho_{\rm   max} = 3.5 \rho_0$. The mean density is $\langle \rho \rangle = 1.535 \rho_0$.

\begin{figure*}
    \plotone{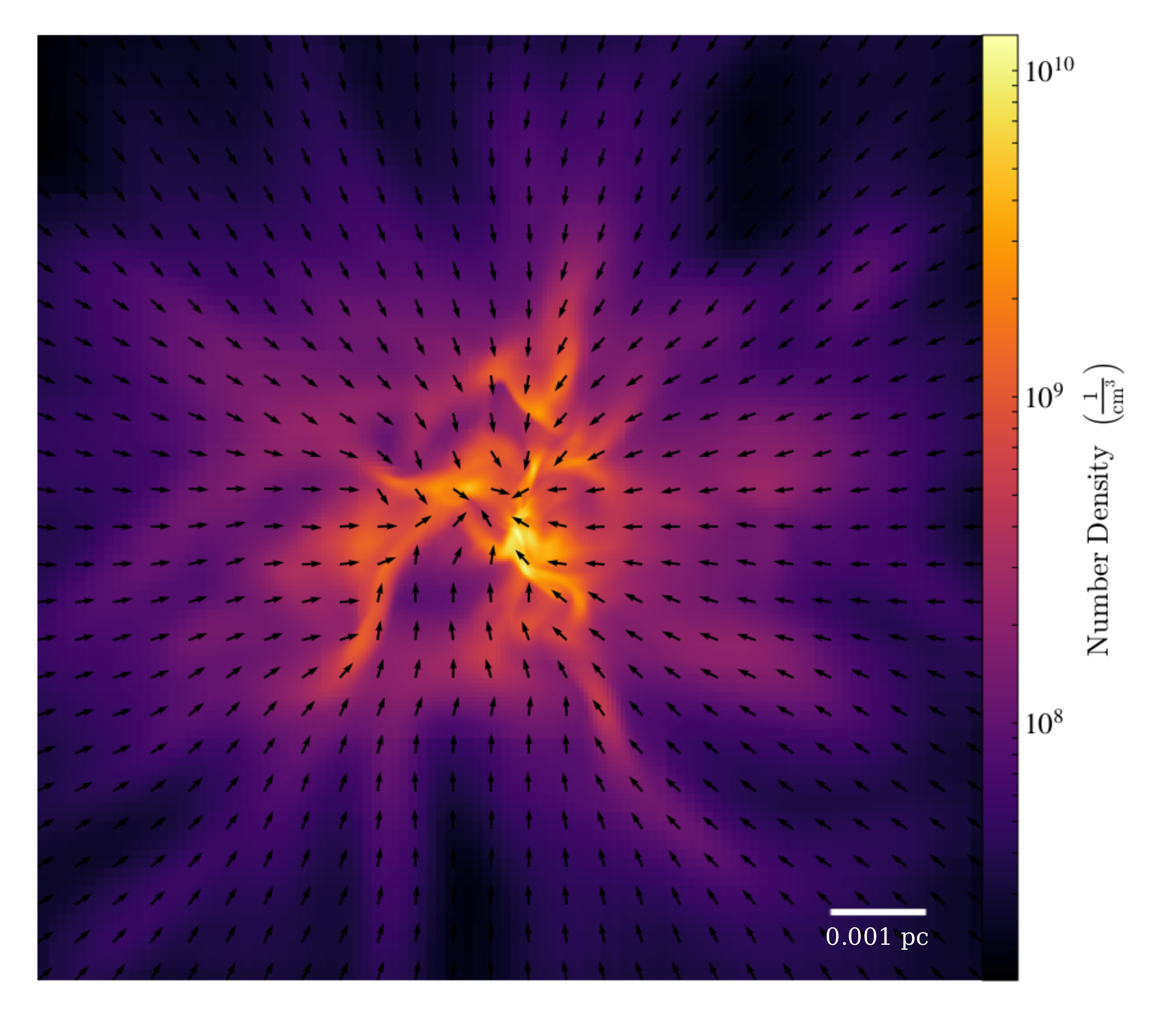}
    \caption{A cross section of the core through its central $(x,y)$ plane at $t \approx \tff$, showing the velocity vectors.} 
    \label{fig:dens_vel_fields}
\end{figure*}

We start the simulations with a turbulent driver to stir the gas, using the prescription of \citet{PF10}, over roughly one crossing time. We then turn off the driving and turn on self-gravity, so that the gas begins to collapse, and the turbulence drains energy from the collapse motions. The initial turbulent velocity dispersion is $\sigtd \approx 0.8 \cs$. The energy is injected in a range of scales between $\Lb/8$ and $\Lb/32$, so that the fluctuations fit within the Gaussian peak. The driving is fully solenoidal. We do not include magnetic fields in our simulation. Table \ref{table:table1} gives a summary of the parameters of the simulations. Figure \ \ref{fig:dens_vel_fields} shows one cross section of the turbulent core at time $t = \tff$. Note that, as pointed out by \citet{Larson69}, the actual collapse time of the simulation is longer than $\tff$ because at the early stages of the collapse, the thermal pressure gradient is not negligible.

\begin{table*}
\begin{center}
\caption{Run parameters.}
\begin{tabular}{lccc}
\hline
\hline
\label{table:table1}
    Run & Effective refinement (pc) & Cells / Jeans length & $\sigtd/c_s$ \\ 
    \hline
    no turb  & $9.76\times10^{-5}$ & 12	& $0.0$         \\
    turb 08  & $9.76\times10^{-5}$ & 12	& $0.8$         \\
    turb 10  & $2.44\times10^{-5}$ & 24	& $0.8$         \\
    \hline
\end{tabular}
\end{center}
\end{table*}

We note that, according to the standard Jeans criterion \citep{Truelove+97}, the local Jeans length should be resolved by at least 4 cells. Actually, we resolve it with a significantly larger number of cells, in order to avoid excessive dissipation within the core. This is because, as is well known \citep[e.g., ] [] {WS85, KC10}, the inner part of a prestellar core, in which the density is roughly flat and the infall speed increases linearly with radius, has a radius of roughly one Jeans length of the central density. Therefore, the resolution applied to the Jeans length is equivalent to that applied to the central part of the core, which we want to be sufficiently resolved as to avoid excessive numerical dissipation of the turbulence within the core. We thus use 12 cells per Jeans length in the low-resolution run and 24 in the high-resolution one.

\section{Results} \label{sec:results}

\subsection{Measurements} \label{sec:measurements}

The presence of the initial Gaussian density profile with the maximum at the center of the simulation box causes the collapse to focus onto this point, and also causes the collapse to adopt an approximately spherical symmetry. In what follows, we analyze only the central part of the core, where the density is nearly flat, and the inwards radial velocity increases linearly with the radius \citep[e.g., ] [] {WS85, Naranjo+15}. Beyond this radius, the velocity begins to decrease again, as also observed in \citet{KC10} and \citet{Naranjo+15}.

Note that in this section we will consider the infall speed of the non-turbulent run as the effective value of $\vg$, the gravitational velocity, rather than the value given by eq.\ (\ref{eq:vg}). This is because, in practice, a marginally unstable isothermal sphere collapses at a rate significantly lower than the free-fall rate, since the thermal pressure is never negligible \citep[see Appendix C of][]{Larson69}, and so the actual infall speed is lower than the free fall value. Thus, in what follows, we will refer to the infall speed of the non-turbulent run simply as the infall (or gravitational) speed denoted by $v_\mathrm{g,sim}$.

To separate the infall component from the turbulent components, we follow the procedure used by \citet{VS+98}, and convert the grid from Cartesian to spherical coordinates to separate the radial ($\vrad$) and the tangential components ($\vphi$ and $\vthet$), which we will assume represents only the turbulent velocity. Then, the tangential velocity satisfies eq.\ (\ref{eq:vtan}), and we assume that it represents the one-dimensional turbulent velocity dispersion; i.e., that
\beq
    \sigod^2 \equiv \langle v_\mathrm{tan}^2\rangle.
    \label{eq:equiv_vtan_sigod}
\eeq
We then average the density as well as the radial and tangential velocity components over spherical shells of thickness equal to one cell, and compute these averages as a function of the radial distance from the center of the collapse.

\begin{figure}[htb] 
    \label{fig:vr_profiles}
    \includegraphics[scale=0.33]{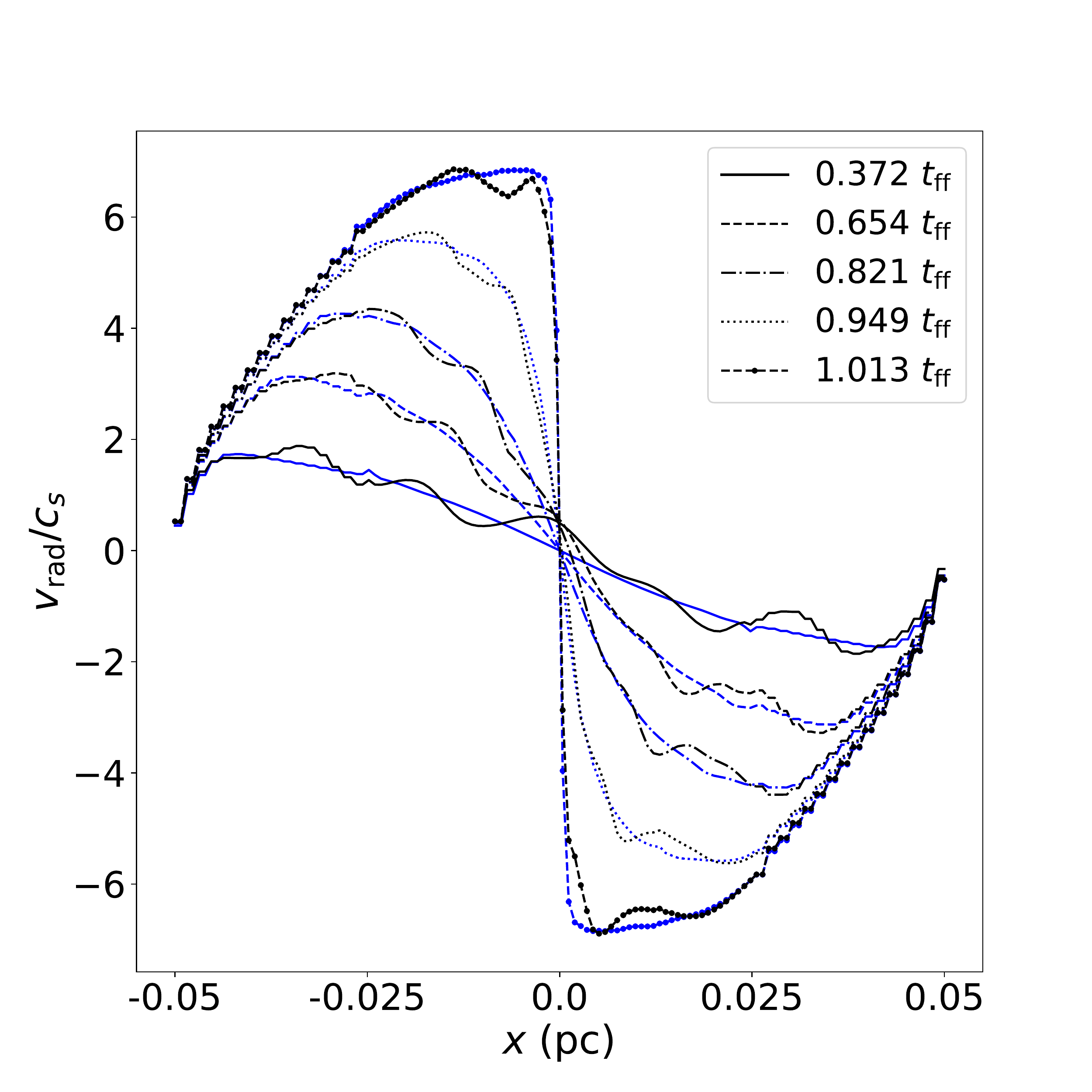}
    \caption{Radial velocity profiles of run turb 08 (black lines) and of the non-turbulent run (blue lines) at various times. Except for the random fluctuations in the turbulent case, the two simulations are seen to have the same mean radial velocity at all times.}
\end{figure}

Figure \ref{fig:vr_profiles} shows the radial profile of the infall velocity at various times for the fiducial turbulent run and for the non-turbulent one.  The inner region where the inwards velocity is approximately linear with radius \citep{WS85} is clearly seen, and is seen to become smaller as time progresses. As shown by those authors, the density profile is nearly flat over this region, whose extent is approximately one Jeans length of the central density \citep{KC10}. The infall speed is thus maximum at the edge of the central flat region.  In what follows, we consider the infall speeds of the simulations at half the radius of this central, flat-density region. In the turbulent simulations, its extent is determined by angle-averaging the radial velocity at each radius. This procedure averages out the turbulent component, and thus only the infall component remains.

To avoid contamination from the boundary conditions, we only consider the radial dependence of the velocities within this region, which radius is resolved at all times by 12 or 24 cells during the contraction, as required by the refinement criterion in the low- and high-resolution simulations, respectively.

An important point to note from Fig.\ \ref{fig:vr_profiles} is that the radial velocity profile of the turbulent run at each timestep is very similar to that of the nonturbulent run, implying that the collapse in the presence of turbulence occurs at the same rate as in the nonturbulent case. We discuss the implications of this result in Sec.\ \ref{sec:implic}.

\begin{figure} 
    \includegraphics[scale=0.33]{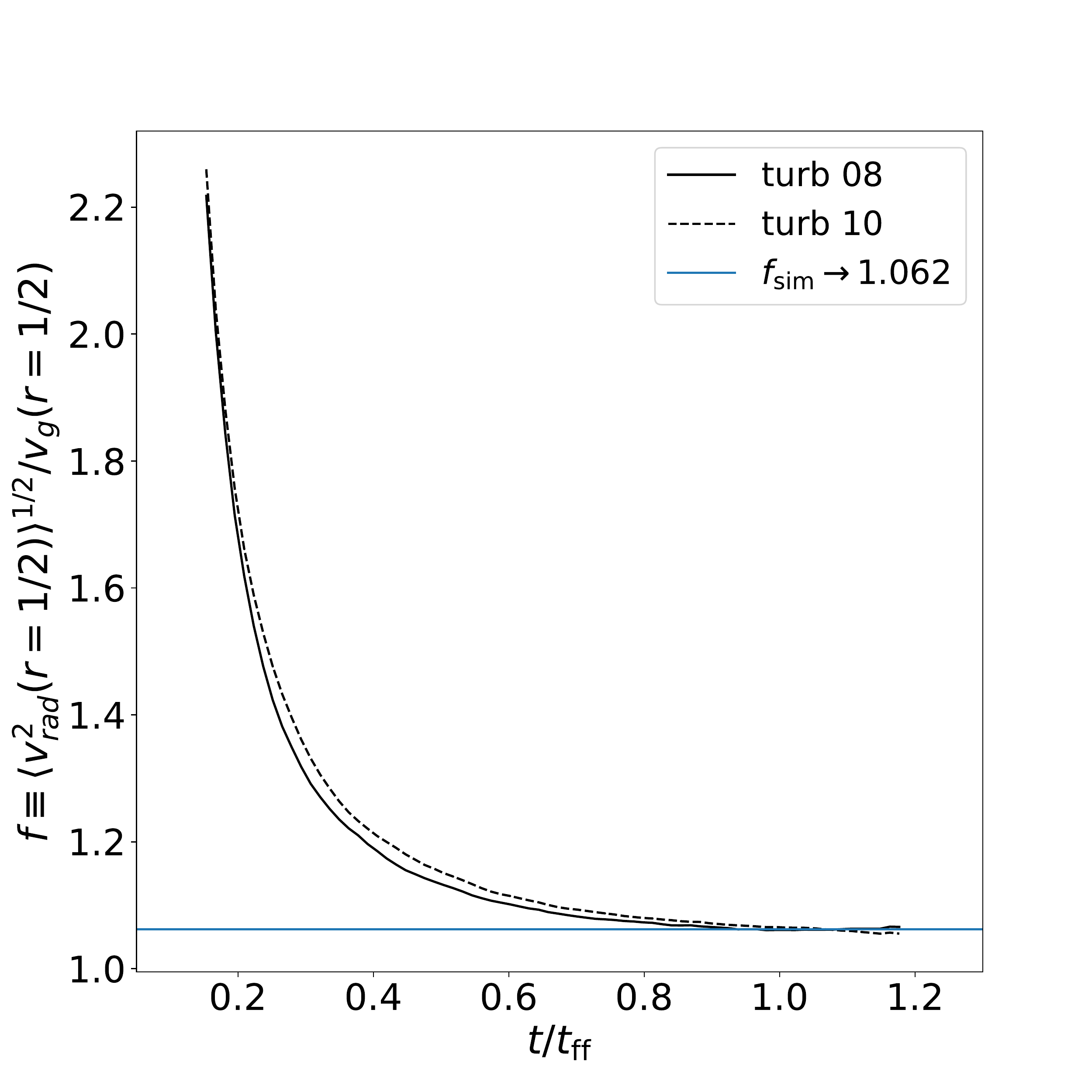}
    \caption{Evolution of the ratio of the radial velocity dispersion (i.e., the average of the squared radial component of the velocity) at half the radius of the flat-density region in the turbulent runs to the dispersion of the infall speed at the same radial position in the non-turbulent run.}
    \label{fig:vrturb_noturb}
\end{figure}

Figure \ref{fig:vrturb_noturb} shows the evolution of the ratio of the average of the radial velocity $\langle v_\mathrm{rad}^2 (r_{1/2})\rangle^{1/2}$ at half the radius of the flat-density region in the turbulent runs to the infall speed $v_g$ at the same position in the non-turbulent run. We observe that, at early times, the ratio is rather large, because the turbulent velocity dispersion is largest, while the infall speed is very small (in fact, it is zero at $t=0$). On the other hand, at larger times, the ratio appears to saturate at a value $f \approx 1.062$.

\begin{figure}[htb]
    \label{fig:vt2_to_vr2}
    \includegraphics[scale=0.33]{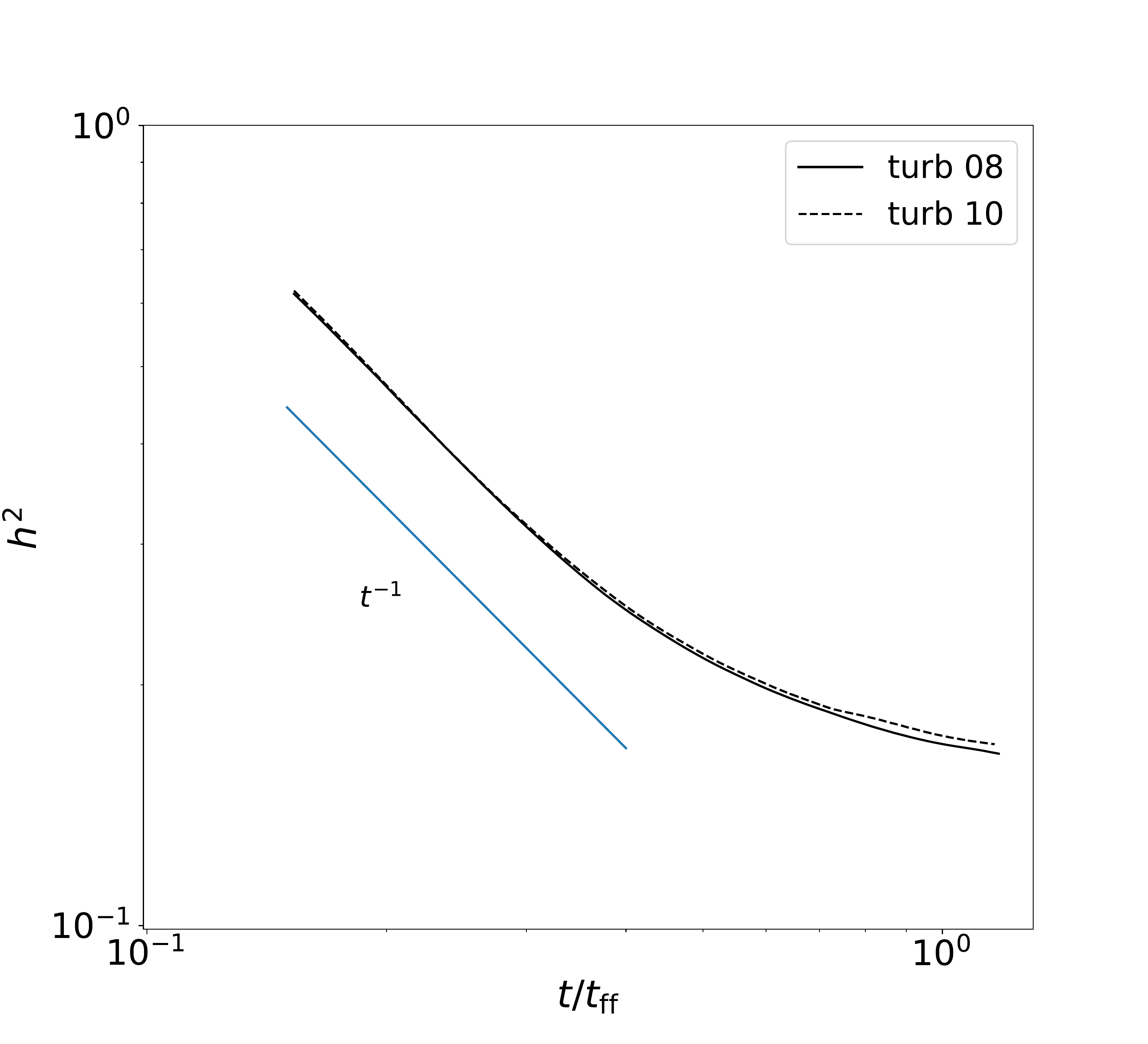}
    \caption{Evolution of the ratio of the 1D turbulent velocity dispersion squared to the angle-averaged radial velocity squared, both measured at half the radius where the maximum infall velocity occurs, for the two turbulent simulations. The ratio first decreases as a power law of time with slope $\sim -1$, and then it is seen to approach a stationary value $\vtan^2/\vrad^2 \approx 0.164$ by $t \approx \tff$.}
\end{figure}

Fig.\ \ref{fig:vt2_to_vr2} shows the evolution of the ratio $\hsim^2 = \vtan^2/\sigvr$ of the square of the tangential velocity to the square of the radial velocity in the simulation, or, equivalently, the ratio of the 1D turbulent velocity dispersion to the total inwards kinetic energy, $\sigod^2/\sigvr$, both measured at half the radius of the flat-density region. It is seen that this ratio first decreases, and then approaches stationarity at a value $\hsim^2 \approx 0.164$.

The evolution of this ratio can be understood considering that the tangential velocity dispersion contains only the turbulent contribution, while the radial velocity dispersion contains this contribution plus that from infall. At the initial condition (not shown in Fig.\ \ref{fig:vt2_to_vr2}), the gravitational velocity is zero, so that $\sigvr = \sigod^2$, and thus $\hsim^2 = 1$. After that, the initial decrease of this ratio indicates that the gravitational speed increases faster than the turbulent velocity dispersion.

Also, the final, nearly stationary value of the ratio $\hsim^2 \approx 0.164$ in Fig.\ \ref{fig:vt2_to_vr2} indicates that, at that level, the energy injection from the infall into the turbulence balances the turbulent dissipation, as considered in Sec.\ \ref{sec:stat_reg}.

From these measurements, we can estimate $g$ in two different ways. On the one hand, from the second equality in eq.\ (\ref{eq:sig_vrad_ratio}), we have 
\beq
    g_{{\rm sim},1} = \sqrt{\fsim^2 \hsim^2} \approx 1.062\, (0.164)^{1/2} \approx 0.43,
    \label{eq:gsim1}
\eeq
on the other hand, we can also determine $\gsim$ by dividing eq.\ (\ref{eq:vdisp_rad}) by $\sigvr$ to obtain
\[
1 = h^2 + \frac{1} {\fsim^2} = \frac{g^2 + 1} {f^2},
\]
so that 
\beq
    g_{{\rm sim},2} = \sqrt{\fsim^2 - 1} \approx 0.36.
    \label{eq:gsim2}
\eeq

We see that the values of $\gsim$ given by eqs.\ (\ref{eq:gsim1}) and (\ref{eq:gsim2}) are similar but not equal. This is probably an indication that our working hypotheses, for example, eqs.\ (\ref{eq:vtan}) and (\ref{eq:vdisp_rad}), apply only approximately. In this case, we can interpret the difference between the two estimates as giving the range of uncertainty of our measurements, and so we write
\beq
    \gsim = 0.395 \pm 0.035,
    \label{eq:gsim}
\eeq

It is noteworthy that this value is very similar to that expected in virial equilibrium (cf.\ eq.\ [\ref{eq:g_vir}]), in spite of the fact that our core is collapsing essentially at the speed of the non-turbulent run. That is, there is no slowing down of the collapse by the turbulence generated, in spite of it being nearly at the virial value. We discuss the possible interpretation of this result in Sec.\ \ref{sec:implic}.

In addition, since the energy ratio attains a nearly stationary value, we can use eq.\ (\ref{eq:g_stat}) to estimate $\Ld/(\eta R) \approx 9.85$ for our system. According to Fig.\ \ref{fig:vr_profiles}, $\langle \vrad \rangle/\vg \approx 1$ by the end of the simulation, and so eq.\ (\ref{eq:g_stat}) reduces to
\beq
    \frac{\Ld} { \eta R} \approx 4 \pi \sqrt{3}\, \gsim \approx 8.60 \pm 0.76.
    \label{eq:g_stat_sim}
\eeq
We discuss the implications and possible interpretations of this result in Sec.\ \ref{sec:inj_scale_dissip_eff}.

\subsection{Convergence study}
\label{sec:convergence}

\begin{figure}[htb]
    \label{fig:figure3}
    \includegraphics[scale=0.33]{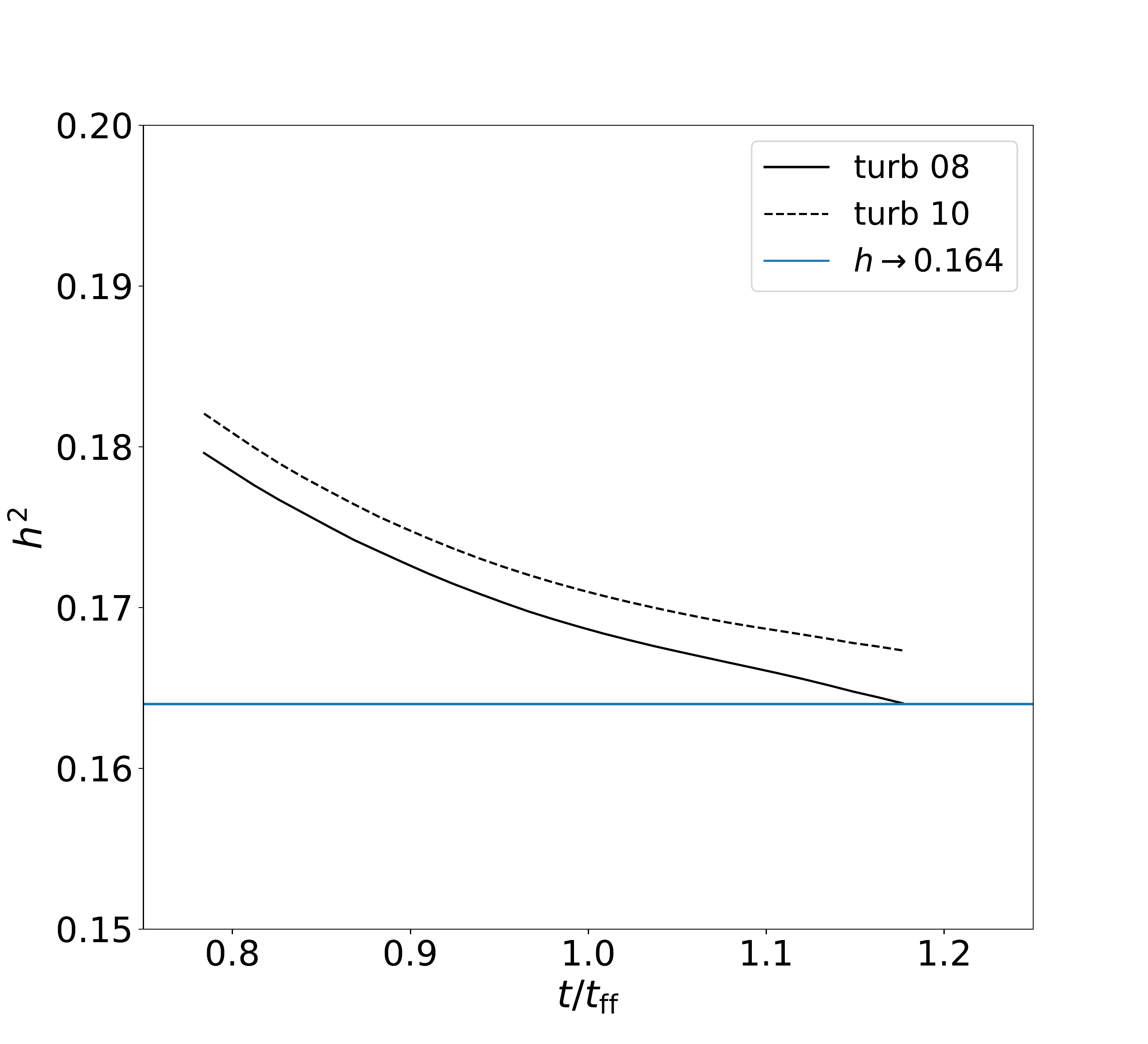}
    \caption{Zoom of the ratio for the final times turbulent over the collapsing kinetic energy through the time. The solid line represents the run with a maximum refinement level of 8, and the dashed line represent the run with a maximum refinement level of 10.}
\end{figure}
In order to test the reliability of our results upon changes in resolution, we performed an additional simulation ``turb 10'', in which we increased the maximum refinement level two levels above the fiducial run ``turb 08'' (cf.\ Table \ref{table:table1}), and the number of cells per Jeans length in the refinement criterion is doubled, to 24. The dashed line in Fig.\ \ref{fig:vrturb_noturb} shows the evolution of the ratio of the root mean square radial velocity $\langle \vrad^2 \rangle$ at the radius where the infall speed is maximum in the high-resolution turbulent run to the same quantity in the non-turbulent simulation. It is seen that this ratio remains within 1\% of the value obtained for the low-resolution run (solid line), implying that our results are very well converged, and that the collapse time is real, rather than an artifact of numerical dissipation.

\section{Discussion}
\label{sec:disc}

\subsection{Implications}
\label{sec:implic}

\subsubsection{The injection scale and the dissipation efficiency}
\label{sec:inj_scale_dissip_eff}

The analytical study of Sec.\ \ref{sec:analyt_pred} provided us with a range of expected values for the velocity ratio $g$ in the cases of energy equipartition (eq.\ [\ref{eq:g_equip}]), virial balance (eq.\ [\ref{eq:g_vir}]), and a self-consistent stationary state (eq.\ [\ref{eq:g_stat}]). In particular, equating eq.\ (\ref{eq:g_stat}) to the value obtained in our simulation, eq.\ (\ref{eq:gsim}), and recalling that $\langle \vrad \rangle /\vg \approx 1$ in our simulation, we find
\beq
    \frac{\Ld} {R} \approx 0.395 \times 4 \sqrt{3} \pi \eta \approx 8.60 \eta.
    \label{eq:gst_eq_gsim}
\eeq

This equation can be interpreted in various ways. We can either assume that $\eta = \etaml$ and determine the implied ratio of $\Ld/R$, or else we can assume that $\Ld = R$ and infer the corresponding value of $\eta$ for our problem. A third option is to assume neither one of the previous possibilities, and take eq.\ (\ref{eq:gst_eq_gsim}) as the final result from our study. For the first two cases, we find:

\medskip
\noindent
{\sc Case I:} $\eta = \etaml$. In this case, we obtain 
\beq
    \Ld \approx 0.57 R,
    \label{eq:eta_eq_etaml}
\eeq
implying that the driving scale is even smaller than the radius, contrary to the assumption made by \citet{Li18}, that $\Ld \sim 2R$.

\medskip
\noindent
{\sc Case II:} $\Ld = R$. In this case, we obtain
\beq
    \etasim \approx 0.12,
    \label{eq:Ld_eq_r}
\eeq
a value $\sim 70\%$ larger than that reported by ML99, $\etaml \approx 0.067$ (cf.\ Sec.\ \ref{sec:stat_reg}).

It is not obvious which one of these two possible interpretations may be more realistic. On the one hand, it is reasonable to assume that the energy injection scale is of the order of the radius, but somewhat smaller, since all points in the core will traverse a distance equal to their respective radial positions by the end of the collapse, and these distances are equal or smaller than the radius of the entire core. On the other hand, it is also reasonable to assume that the dissipation efficiency will be of the order of that found by ML99, but not quite equal, since the nature of the driving mechanism, the geometry, and the numerical codes are different. All in all, probably the most appropriate conclusion is that the injection scale is of the order of the radius, and that the dissipation efficiency is of the same order as that reported by ML99.

\subsubsection{The non-adiabatic nature of the turbulence driving by collapse} 
\label{sec:thermodynamics}

The generation of turbulence by gravitational collapse is often referred to as an ``adiabatic heating'' of the turbulence \citep[e.g.,] [] {RG12, MC15, Li18, Xu+20, Mandal+20}. Although this terminology arises from the fact that, in the First Law of Thermodynamics, the $P\, dV$ work corresponds to an adiabatic heating process, it is important to keep in mind that turbulence is inherently a dissipative process, and so the net driving (or ``heating'') of the turbulence is far from being adiabatic, a term which, by definition, implies that no loss of heat or mass occurs during the process. In a standard thermodynamic system, this would mean $dQ=0$. In the case of turbulence driving (or ``heating''), this would mean no turbulence dissipation. However, in this case, the dissipative loss of turbulent kinetic energy is unavoidable since, contrary to the case of microscopic molecular motions, in which particle collisions are ellastic, the ``collision'' of gas streams produces vortices and shocks where the kinetic energy is intrinsically converted to heat. This is in fact at the heart of the theory of \citet{K41}, in which the slope of the turbulent energy spectrum is determined by the condition that the dissipation rate equals the injection rate and the energy transfer rate among scales. 

The results by \citet{RG12} and \citet{Xu+20} suggest that, when the collapse rate is larger than the turbulent eddy turnover rate, then the turbulence {\it gains} kinetic energy, and this is often referred to as ``adiabatic heating of the turbulence''. In the opposite case, when the turbulent turnover rate is larger than the collapse rate, the turbulent {\it loses} energy. Nevertheless, even in the case when the turbulence gains energy, dissipation is still active, and thus the term ``adiabatic'' is inadequate. A more appropriate term should be ``net heating.''

The non-adiabatic nature of the turbulence generation by the collapse is manifested by the fact that the collapse does not appear to be significantly delayed by the turbulence, even though a nearly virial turbulent level is attained during the collapse. This can only be understood if the turbulent energy is dissipated as quickly as it is produced by the gravitational contraction, preventing it from delaying the collapse, contrary to the case of truly nearly adiabatic heating of a collapsing protostellar object when it becomes optically thick, trapping the heat, and producing a first hydrostatic core. Instead, our simulations show that no such halting, or even delaying, of the collapse is accomplished by the turbulence even when it appears virialized, fed by the collapse itself. This further implies that, in particular, when writing the virial theorem for a collapsing object including a nonthermal kinetic energy term \citep[e.g.,] [] {MZ92, BP+99, BP06, MO07}, an additional term must be included to represent viscous dissipation of the turbulent energy. 

The standard procedure for deriving the virial theorem is to calculate the work done by the forces by dotting the momentum equation with the position vector and integrating over a volume \citep[see, e.g.,] [] {Shu92}. For our non-magnetic case, the specific momentum equation can be written as
\begin{equation}
    \frac{d\mathbf{u}}{dt} = - \frac{\partial P}{\rho} - \nabla\varphi + \nu\nabla^2\mathbf{u},
\end{equation}
where, for simplicity, we have written the viscous term appropriate for incompressible flows. Since this is only for illustrative purposes, it does not make a significant difference, as the compressible terms also consist of second-spatial derivatives of the velocity, of the form $\nabla \nabla\cdot \mathbf{u}$. Also, considering a cold cloud, we can neglect the thermal pressure, and obtain the following expression for the virial theorem
\begin{equation}
    \frac{d^2I}{dt^2} - 2K = W + \mathcal{D},
\end{equation}
where $I = \int x^2 \rho dV$ is the moment of inertia, $K = \int \rho u^2 dV$ is the kinetic energy associated to the non-thermal velocity dispersion, $W$ is the gravitational energy, and $\mathcal{D}$ is the work done by the viscosity, defined as $\mathcal{D}=\int_V\nu\mathbf{x}\cdot\nabla^2\mathbf{u}dV$. It is important to note that there is no virial equilibrium ($d^2I/dt^2=0$) if the core is in gravitational collapse. This means that the fact that we observe $2K \approx W$ does not imply that the system is in equilibrium, because in this case there are two other terms on the virial theorem that inter the energy budget. We plan to investigate this issue in a future contribution.

\begin{figure}[htb]
    \label{fig:PtPr}
    \includegraphics[scale=0.33]{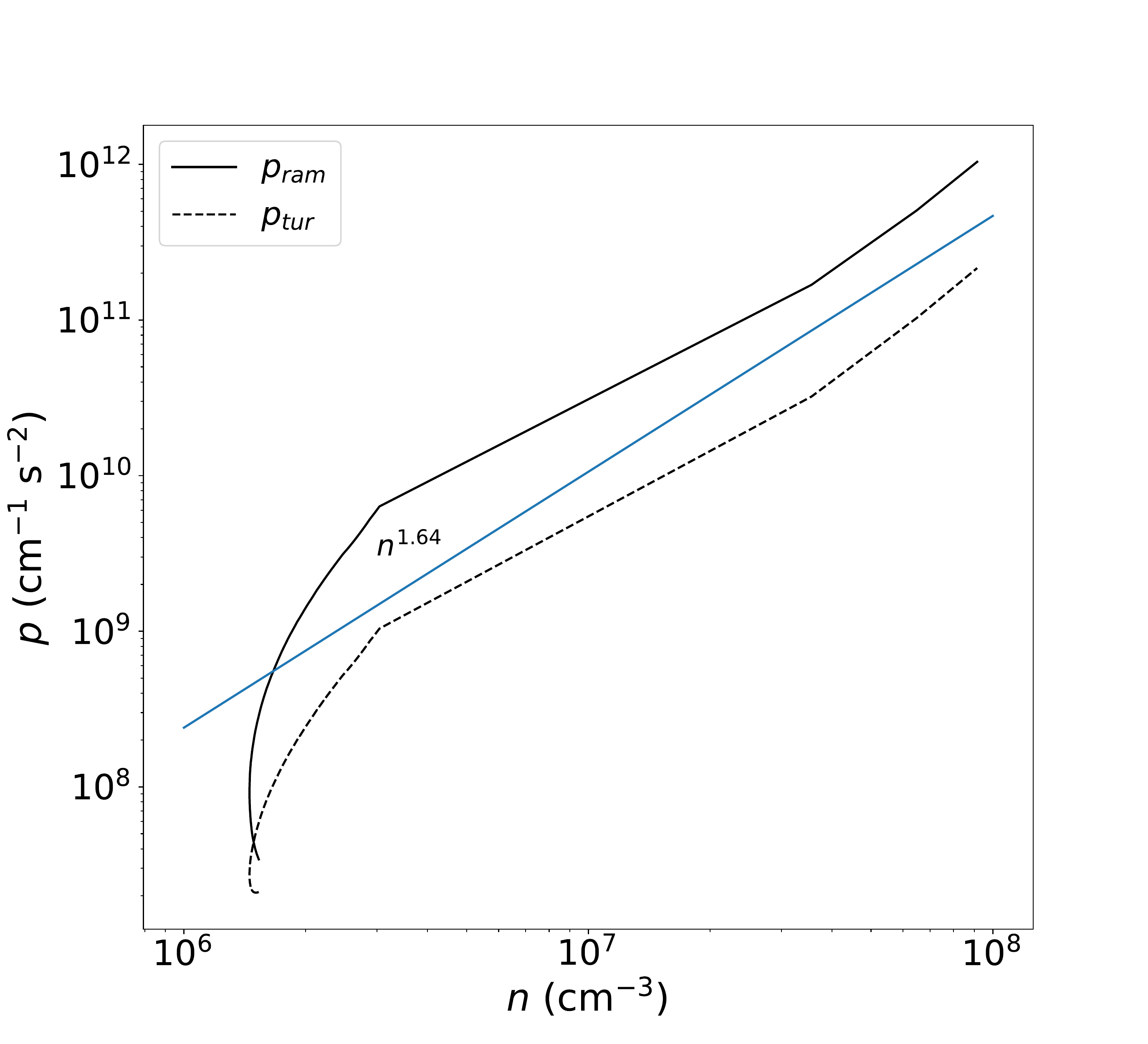}
    \caption{The solid line represents the ram pressure, the dashed line represents the turbulent pressure, and the blue line is a fit as a power law with a slope of 1.64.}
\end{figure}

The lossy character of the turbulent ``heating'' can also be seen in the effective polytropic exponent displayed by the turbulent and infall ram pressures, respectively defined as $\Pt \equiv \rho \sigod^2$ and $\Pram\equiv \rho \vrad^2$. The solid and dashed lines in Fig.\ \ref{fig:PtPr} show these pressures, averaged over spherical shells, {\it versus} the corresponding average density in the shells. The blue line shows a slope of $\gamef \sim 1.64$, which is a fit to the slope of these curves. This result is consistent with the study by \citet{VS+98} who, using a spectral code to simulate a turbulent gravitational collapse, found values of the polytropic exponent $\gamef \approx 3/2$ for nonmagnetic and rapidly collapsing magnetized cases, and $\gamef \approx 2$ for slowly collapsing, strongly magnetized cases.

Also, it is noteworthy that this slope is very similar to the value of the polytropic exponent corresponding to an adiabatic process in a monoatomic gas, $\gamma = 5/3$, which in particular is larger than the well known critical value of $4/3$ required for the internal thermal energy to be capable of halting gravitational collapse \citep[e.g.,] [] {Chandra61, VS+96}.  Nevertheless, the turbulence turns out to be incapable of storing the energy of the collapse and thus delaying it to any significant extent.  The solution lies in the observation that the ram pressure is always larger than the turbulent one, showing that the turbulent energy, although continuously increasing, is always lagging behind the infall kinetic energy, a result which can only be attributed to the rapid dissipation and to the establishment of equality between the rates of collapse and of turbulent energy transfer. The fact that the result is preserved at higher resolution suggests that this is a real effect and not just an effect of excessive numerical dissipation.

\subsubsection{Comparison with previous work}  \label{sec:comparison}

The problem of turbulence driving (``heating'') by gravitational collapse has been addressed by a number of authors using both analytical and numerical approaches. Numerically, \citet{VS+98} used a very similar approach to the one used here, except with a fixed-grid spectral code, to investigate the effective equation of state of the turbulence. In particular, \citet{VS+98} were searching for a ``logatropic'' behavior \citep{Lizano+89} of the turbulent pressure upon the compression produced by the collapse, but instead found a polytropic behavior with an effective exponent in the range 3/2--2 depending on the collapse regime. The exponent we find here (Fig.\ \ref{fig:PtPr}) is fully within the range determined by those authors. 

More recently, other studies \citep{RG12, Mandal+20} have used numerical simulations of turbulent boxes using shrinking comoving coordinates, to investigate the rate of turbulent generation during contraction. These simulations, in which the contraction rate is a fixed parameter or function, have provided valuable insight on the driving of the turbulence by the contraction, such as the approach of the turbulent eddy turnover rate to the contraction rate \citep{RG12}. Similarly, using a thermally bistable gas, \citet{Mandal+20} found that the physical properties of the dense clouds in their simulations best matched those of real molecular clouds when the two rates are equal, again pointing towards a balance between the two rates in molecular clouds.

Our finding that the turbulence approaches a ``virial'' value is in qualitative agreement with their result. However, because the simulations of \citet{RG12} and \citet{Mandal+20} have a fixed  contraction rate, they cannot determine self-consistently whether the collapse can be delayed or not by the turbulent pressure. 

On the other hand, in our simulation, the ratio of the infall energy to the turbulent energy self-consistently increases over time as the collapse evolves, because the initial contraction rate is zero, and then it increases and overtakes the turbulent rate. The latter was nonzero from the start, but decayed until the collapse rate was large enough to produce sufficient driving of the turbulence. Therefore, our simulation effectively describes a trajectory in the parameter space of the collapse rate to turbulent rate ratio. This evolution results from the self-consistent treatment of the collapse and the effect of the turbulence. With this self-consistent prescription, we find that the turbulence generated by the collapse at no point is able to delay the collapse.

From the analytical standpoint, various workers have considered an energy balance equation for the energy injection from the contraction and the turbulent energy dissipation to compute the turbulent velocity dispersion, and from there compute a turbulent pressure which is added to the regular thermal pressure in the momentum equation for the infall speed \citep{RG12, MC15, Li18, Xu+20}, with various results. As mentioned above, \citet{RG12} concluded that the turbulent transfer rate must approach the contraction rate. On the other hand, \citet{MC15} consider a {\it protostellar} (or post-singularity) core, in which a central star or cluster has already formed, and define two distinct radial regions, separated by the radius that demarks the sphere of influence of the mass accreted onto the central object. They then conclude that ``turbulent pressure'' is important at all radii, and that the infall speed is smaller (resp.  comparable) than the turbulent velocity outside (resp.\ inside) that radius. Since our simulations are restricted to the {\it prestellar} stage (i.e., pre-singularity), we cannot assess whether our simulations support their analytical result. 

In addition, \citet[] [hereafter XL20] {Xu+20} have performed a similarity analysis of the ``inside-out'' collapse problem \citep{Shu77}, in which they consider a fixed ratio $C$ of the turbulent velocity (equivalent to our $\vtan$ and $\sigod$) to the total radial velocity (equivalent to our $\vrad$), so their $C$ is equivalent to our $h$.  Under the assumption of a constant $C$, XL20 consider cases with different initial values of the ratio of the infall speed to the sound speed, and find that, when this ratio is initially large, significant (supersonic) turbulent energy is generated by the collapse, and that this turbulence has the effect of modifying the radial density and velocity profiles. The resulting infall velocity is uniform at large radii, it decreases inwards at intermediate radii, and then increases inwards at small radii. This infall velocity profile is actually somewhat similar to the classical prestellar profile, which has uniform velocity at large radii, and a linearly decreasing velocity at small radii \citep[] [see also Fig.\ \ref{fig:vr_profiles}] {WS85}, with the difference that the velocity increases at small distance from the center. From these results, XL20 conclude that the turbulent pressure can slow down the collapse in the case of an initially large infall speed.  

However, the general approach of considering the generation and dissipation of turbulence by the collapse and then feeding it back via a turbulent pressure term in the momentum equation has a number of important caveats, as we now discuss. 

First, this approach implicitly assumes that the turbulent motions occur at such a small scale that they can be considered analogous to the thermal velocity dispersion of the gas molecules, neglecting the fact that the largest turbulent speeds occur at the largest scales within the system. Therefore, the effect of the turbulence is not to act as an isotropic pressure, but rather as a distorting agent for the density structures \citep{BP+99}. As a consequence, rather than slowing down a monolithic collapse, large turbulent velocities (larger than the infall speed) are expected to cause fragmentation, shearing or compression of the collapsing core. XL20 explicitly state that, contrary to the case of turbulence driven by external compressions, for collapse-generated turbulence, the driving scale $\Ld$ is small. However, our result that $\Ld \la R$ implies that it is still of the order of the radius of the collapsing core itself, so it is not small compared to the system size. On the other hand, it is not as large as the core's diameter, in contradiction to the {\it ad hoc} assumption of \citet{Li18}, which was important for the conclusion by this author that the collapse can be retarded by the turbulence. 

Within the framework of turbulent pressure support, both \citet{MC15} and XL20 argue that its effect is to modify the velocity dispersion-size relation from the standard \citet{Larson81} form, $\sigma \propto R^{1/2}$, to scalings with significantly smaller exponents $\sim$ 0.2 -- 0.3. Both groups propose this as an explanation of the observation that dense massive cores often exhibit flatter scalings than Larson's \citep[e.g.,] [as plotted by \citealt{BP+11}] {Caselli+95, Plume+97, Shirley+03, Gibson+09, Wu+10}. However, both \citet{MC15} and XL20 have neglected the fact that those massive cores do follow the appropriate gravitational scaling once the additional dependence on the column density \citep[$\sigma \propto \sqrt{\Sigma R}$;] [] {Keto+86, Heyer+09}, is taken into account, as shown by \citeauthor{BP+11} (\citeyear{BP+11}, see also \citealt{Camacho+16}; \citealt{BP+18}). Those, the apparent deviations from Larson's linewidth-size relation can be fully accounted for by gravitational contraction when the column density dependence is included, with no need for turbulent support. 

Second, in the similarity studies, the ratio of the contraction rate to the turbulent transfer rate is a parameter defined as an initial condition, because, by their very nature, similarity methods may not be consistent with the initial and boundary conditions of the problem \citep[e.g.,] [Ch.\ 17] {Shu92}, and the solutions only apply once the initial transients have passed. However, it is precisely those initial transients that determine in a self-consistent manner the parameters of the similarity treatments. In this sense, our numerical simulations do evolve self-consistently from a much earlier stage than the post-protostar-formation stages considered by \citet{Shu77}, \citet{MC15} and XL20. Note, however, that our simulations do eventually develop a regime consistent with the similarity solutions corresponding to the prestellar collapse \citep{WS85} after the initial transients have passed.

In particular, our simulations have a period during which our $h$ parameter (equivalent to XL20's $C$ parameter) is not constant, but evolves towards a final nearly stationary value. Indeed, our turbulent simulations seem to evolve towards $h \sim \sqrt{0.164} \approx 0.405$ (cf.\ Sec.\ \ref{sec:measurements}).
At those stages of our {\it prestellar} collapse, the infall speed is large compared to the sound speed (cf.\ Fig.\ \ref{fig:vr_profiles}) and there has been significant turbulent generation (as indicated by its nearly virial value; cf.\ eq.\ [\ref{eq:gsim}] and the subsequent discussion),  in agreement with Case 2 of XL20. Nevertheless, we do not observe any significant retardation of the collapse at any time nor radius. We speculate this is because our simulation evolves towards equality of the injection and dissipation rates, in which case XL20 find that the infall is undisturbed.  

\subsection{Caveats} \label{sec:caveats}

Our work is certainly limited by some caveats. First, we have chosen to use an adaptive-mesh scheme in order to self-consistently be able to follow the collapse, in particular whether it is delayed by the turbulence generated. This cannot be accomplished using a fixed-resolution box with shrinking coordinates, as done by \citet{RG12} and \citet{Mandal+20}, since in that case, the contraction rate is imposed as an external parameter. However, our choice implies that, at late stages, our core is only moderately resolved, possibly generating excessive numerical dissipation. Our high-resolution simulation, with twice the resolution in the core,  still collapses in essentially the same time, and converges to nearly the same values of the velocity ratios, suggesting our result is robust. Nevertheless, it would be desirable to perform a test at much higher resolution, which we will attempt in a future study.

Second, we have chosen to use a very low level of initial turbulence in order to avoid the destruction of the core by the initial turbulence and thus approximately maintain the global spherical symmetry. It could be argued that this is the reason the collapse is not delayed. However, the fact that by the end of the turbulent simulations the turbulence has reached a nearly virial value, suggests that at those late stages the turbulence should be able to slow down the collapse. The fact that is does not strongly indicates that it is dissipated as rapidly as it is generated, which is in fact what we observe: a stationary balance between injection and dissipation.

Third, we are not taking into account the role of the magnetic field. The role of the magnetic field in the collapsing core problem is likely to be important in the efficiency of turbulence driving by the collapse, although it may not be the dominant component, since it is now established that most molecular clouds and their cores are magnetically supercritical \citep{Crutcher12}. But even in this case, magnetic fields may have two opposite effects in the efficiency of turbulence generation. On the one hand, the magnetic field could propagate the fluctuations via Alfvén waves, aiding the transfer of turbulent kinetic energy throughout the core. On the other hand, if the magnetic field is strong enough, it may restrict the velocity field to be oriented preferentially along field lines, inhibiting transverse motions. Since the collapse causes an hourglass (nearly radial) morphology of the field lines, the tangential motions in our core might be inhibited, thus inhibiting the transfer of energy from the radial collapse direction to the random turbulence. We plan to investigate the competition between these effects in a forthcoming study.

Finally, our simulations have been limited to the prestellar (pre-singularity formation) stage of the collapse, and so they are not directly comparable to some of the studies presented by some other groups, which consider the protostellar stage, with a finite-mass central point object already present \citep[e.g.,] [] {MC15}. However, since the prestellar stage sets the initial conditions for the formation of a protostar, we consider that understanding this stage is of utmost importance. In any case, in a future study we will attempt to follow the entire evolution from the prestellar to the protostellar stage.

\section{Summary and Conclusions}
\label{sec:concls}

In the present paper, we have performed some simple analytical calculations predicting the ratio $g$ of the one-dimensional turbulent velocity dispersion $\sigod$ to the gravitational velocity $\vg$ under the conditions of free-fall (or energy equipartition, $\Eturb = |\Eg|$), virial balance ($2\Eturb = |\Eg|$) and a stationary balance between energy injection by the collapse and energy dissipation by viscosity. We then performed a suite of three numerical simulations of spherical collapse, two in the presence of initial turbulence at different resolutions, and one with no initial turbulence. We decomposed the velocity in radial and tangential components, identifying the latter with the purely turbulent one dimensional velocity dispersion $\sigod$. We then measured the ratio $f$ as well as the ratio $h$ of $\sigod$ to the mean square radial velocity $\langle \vrad^2 \rangle^{1/2}$, and from there, we inferred the ratio $g$ in the turbulent simulations by two different means. 

In spite of our turbulence having a very small initial amplitude, we found that, before one free-fall time, the ratio $g = \sigod/\vg$ has reached a nearly virial value $\gsim = 0.395 \pm 0.035$. Nevertheless, we also found that $\sigod$ approaches a fixed fraction of the mean squared radial velocity, indicating a stationary balance between injection and dissipation. From here, we inferred the value of the dissipation efficiency $\eta$ or the ratio of the energy injection scale $\Ld$ to the core's radius, finding that $\eta$ is within 70\% of the value reported by ML99, and that $\Ld \lesssim R$. Finally, we measured the effective polytropic exponent of the turbulent ``pressure'', finding a value $\gamef \approx 1.64$, which, at face value, would suggest a nearly adiabatic character of the turbulent pressure upon compression.

Most important, however, is the fact that we did not find any significant slowing down of the collapse in neither of the turbulent simulations, compared to that of the non-turbulent simulation, in spite of $\sigod$ having a nearly virial value, and $\gamef$ being larger than the critical value of 4/3, above which a polytropic gas is capable of halting the collapse. This led us to conclude that the turbulence generated by the collapse is dissipated so rapidly (independently of resolution) that it is unable to delay the collapse at any significant rate. This implies that neither the critical value $\gamef = 4/3$ nor the presence of a nearly virial velocity dispersion can be taken as indicative of delay or halting of the collapse in the presence of strong dissipation, which is equivalent to a loss of heat in a thermodynamic system. Therefore, the ``heating'' of turbulence by gravitational collapse is lossy rather than adiabatic, and that the dissipation of turbulence occurs at the same rate as the injection, implying that it cannot be stored in the system to slow down the collapse.

\acknowledgments
This work has been supported in part by a CONACYT graduate fellowship for R.G.-G.\ and CONACYT grant 255295 to E.V.-S.

\software{FLASH}

\bibliographystyle{yahapj}

\end{document}